\documentclass[]{iopart}
\usepackage{iopams,graphicx}
\usepackage[british]{babel}
\newcommand{\openone}{{\mathchoice {\rm 1\mskip-4mu l} {\rm 1\mskip-4mu l}
{\rm 1\mskip-4.5mu l} {\rm 1\mskip-5mu l}}}
\begin{document}

\title[Quantum stochastic models of two-level atoms]{A quantum stochastic
approach to the spectrum of a two-level atom}

\author{Alberto Barchielli\footnote{Also: Istituto Nazionale di Fisica Nucleare,
Sezione di Milano.}\ and Nicola Pero}
\address{Dipartimento di Matematica, Politecnico di Milano, \\
Piazza Leonardo da Vinci 32, I-20133 Milano, Italy}

\begin{abstract}
By means of quantum stochastic calculus we construct a model for an atom with
two degenerate levels and stimulated by a laser and we compute its fluorescence
spectrum; let us stress that, once the model for the unitary atom-field
dynamics has been given, then the spectrum is computed without further
approximations. If only the absorption/emission term is included in the
interaction, we reobtain the Mollow spectrum in the case of a monochromatic
laser and the Kimble-Mandel spectrum in the case of a ``phase diffusion model''
for a non monochromatic laser. However, our model can describe also another
type of light scattering, a ``direct scattering'' due to the response of the
atom as a whole, which we expect to be small, but which interferes with the
scattering due to the absorption/emission channel. When both the scattering
channels are introduced we obtain a modification of the Mollow-Kimble-Mandel
spectrum, which shares the main features with the usual case, but which
presents some asymmetries even in the case of no detuning.
\end{abstract}


\section{Introduction}\label{intro}

Quantum stochastic calculus (QSC) \cite{1HP84}--\cite{Partha92} is a
noncommutative analogue of the classical Ito's calculus and it has had many
applications in modelling quantum optical systems \cite{Gar86}--\cite{BPag}. In
particular, inside QSC a Schr\"odinger equation for unitary operators can be
formulated, the Hudson-Parthasarathy equation \eref{1.13} \cite{1HP84}, which
is well suited to treat the interaction between matter and electromagnetic
field within the  usual approximation schemes of quantum optics \cite{Bar90}.

The simplest, non trivial matter-field system is a two-level atom
stimulated by a laser. The fluorescence spectrum of such a system,
in the case of a perfectly monochromatic laser, was obtained by
Mollow \cite{Moll}; then, the case of a laser with a Lorentzian
spectrum was developed by Kimble and Mandel
\cite{kim-man1,kim-man2}. QSC and Hudson-Parthasarathy equation
give not only the way of giving an unified treatment of such a
system and of connecting it to the photon-detection theory, but
they allow  also to explore possible corrections to the dynamics
and to obtain modifications of the Mollow spectrum. In the usual
treatments, the scattering of the light by the atom is described
by the absorption/emission process due to the electric dipole
interaction involving only two states of the atom. But even if the
absorption/emission process would be forbidden and the atom would
be frozen in the up or down level, some scattering of light would
remain, due to the response of the atom as a whole; we can call it
``direct scattering''. For instance, in a perturbative development
in Feynman graphs, scattering processes would be generated also by
virtual transitions starting and ending in one of the two states
left in the description, but involving as intermediate states the
other ones, which have been eliminated in the final description.
The use of Hudson-Parthasarathy equation allows, even when the
atom is approximated by a two-level system, for the introduction
of both the ``absorption/emission channel'' and the ``direct
scattering channel''. This modified model, in the case of a
perfectly monochromatic laser, has been presented in \cite{BarL98}
and developed in \cite{BarL2000}; the first results for a non
monochromatic laser have been obtained in \cite{Pero}.

In Section \ref{model} we give the mathematical model for the
dynamics of the two-level atom interacting with the
electromagnetic field, including the polarization of the light and
the degeneracy of the up and down levels, which has never been
considered before and which has an effect on the angular
dependence of the spectrum. We consider the usual dipole
interaction and the possible corrections giving the direct
scattering; the so called phase diffusion model is used for the
laser as in \cite{kim-man1}. In Section \ref{flspectrum} the
general formulae for the spectrum are obtained and its angular
dependence is discussed. Then, the total spectrum, integrated over
the whole solid angle, is studied. In the case of only the
absorption/emission process in the interaction term, we reobtain
the Mollow spectrum \cite{Moll} for a perfectly monochromatic
laser and the Kimble-Mandel spectrum \cite{kim-man1} for a non
monochromatic laser with Lorentzian spectrum. In the case of the
modified model, when also the direct scattering is included, but
the laser is perfectly monochromatic, we reobtain the results of
\cite{BarL2000}, although there the level degeneracy was not taken
into account. According to the values of the various parameters,
the fluorescence spectrum can present a well resolved triplet
structure as in the Mollow spectrum, but also it can be distorted
and made asymmetric by the presence of the new terms. Experiments
\cite{Eze1}--\cite{Walther2} confirm essentially the triplet
structure; some asymmetry has been found and its origin has been
attributed to various causes. The presence of the new terms we
have introduced could give a model for treating such an asymmetry.
When the laser bandwidth is taken into account, the modified model
gives corrections to the spectrum found in \cite{kim-man1} by
Kimble and Mandel.

\section{The model}\label{model}

\subsection{The unitary atom-field dynamics}

\subsubsection{Fock space and the Hudson-Parthasarathy equation.}
Let us recall the rules of QSC and the Hudson-Parthasarathy
equation; this is just to fix our notations, while for the proper
mathematical presentation we refer to the book by Parthasarathy
\cite{Partha92}.

We denote by ${\cal F}$ the Boson Fock space over the ``one-particle space''
$\mathcal{Z} \otimes L^2(\mathbb{R}_+)$, where $\mathcal{Z}$ is a separable
complex Hilbert space; we shall see in section \ref{photonspace} how to choose
$\mathcal Z$. A vector $f$ in $\mathcal{Z} \otimes L^2(\mathbb{R}_+) \simeq
L^2({\mathbb R}_+; {\cal Z})$ is a square integrable function from ${\mathbb
R}_+$ into $\cal Z$. Let $\{h_k,\ k\geq 1\}$ be a c.o.n.s.\ in $\cal Z$ and let
us denote by $A_k(t)$, $A^\dagger_k(t)$, $\Lambda_{kl}(t)$ the annihilation,
creation and gauge (or number) processes associated with such a c.o.n.s.; we
can write formally $A_k(t) = \int_0^t a_k(s) \,{\mathrm d} s$, $ A_k^\dagger(t)
= \int_0^t a_k^\dagger(s) \,{\mathrm d} s$, $\Lambda_{kl}(t) = \int_0^t
a_k^\dagger(s)a_l(s) \,{\mathrm d} s$, where $a_k(t)$, $a_k^\dagger(t)$ are
usual Bose fields, satisfying the canonical commutation rules
$[a_k(t),a^\dagger_l(s)]= \delta_{kl}\delta(t-s), \ldots$. The Fock space
$\mathcal F$ is spanned by the coherent vectors $e(f)$, $f \in L^2({\mathbb
R}_+; {\cal Z})$, whose components in the $0,1,\ldots,n,\ldots$ particle spaces
are
\begin{equation}\label{1.2}
e(f) = \exp\left(-\case{1}{2}\|f\|^2\right)\left(1,f,(2!)^{-1/2}f\otimes
f,\ldots,(n!)^{-1/2} f^{ \otimes n}, \ldots \right);
\end{equation}
in particular the vector $e(0)$ is the Fock vacuum. The coherent vectors
\eref{1.2} are normalized and they are the eigenstates of the annihilation
operators; indeed, we have $A_k(t) e(f) = \int _0^t \langle h_k|
f(s)\rangle\,\rmd s\, e(f)$. In QSC integrals of ``Ito type'' with respect to
${\rmd} A_k(t)$, ${\rmd} A_k^\dagger(t)$, ${\rmd} \Lambda_{kl}(t)$ are defined.
The main practical rules to manipulate ``Ito differentials'' are the facts that
${\rmd} A_k(t)$, ${\rmd} A_k^\dagger(t)$, ${\rmd} \Lambda_{kl}(t)$ commute with
anything containing the fields only up to time $t$ (objects called adapted
processes) and that the products of the fundamental differentials satisfy $
\rmd A_k(t)\, {\rmd} A_l^{\dagger}(t) = \delta_{kl}\, {\rmd} t$,  ${\rmd}
A_k(t)\, {\rmd} \Lambda_{rl}(t) = \delta_{kr}\, {\rmd}
 A_l(t)$,
$
\rmd\Lambda_{kr}(t)\, {\rmd} A_l^\dagger(t) = \delta_{rl}\,
 {\rmd} A_k^\dagger(t)$,
${\rmd} \Lambda_{kr}(t)\, {\rmd} \Lambda_{sl}(t) = \delta_{rs}\, {\rmd}
\Lambda_{kl}(t)$; all the other products and the products involving $\rmd t$
vanish.

Let ${\cal H}$ be a separable complex Hilbert space (the system space) and let
$R_k$, $k\geq 1$, $S_{kl}$, $k,l\geq 1$, $H$ be bounded operators in ${\cal H}$
such that $H^\dagger =H$, $\sum_k R_k^{\dagger } R_k$ is strongly convergent to
a bounded operator, and
\begin{equation}
\sum_{kl} S_{kl} \otimes |h_k \rangle \langle h_l|=: S\in {\cal U}({\cal
H}\otimes {\cal Z}),
\end{equation}
$\mathcal{U}( *)$ ``means unitary operators on $*$''; we set also
\begin{equation}
R : \mathcal{H} \to \mathcal{H} \otimes \mathcal{Z}\,, \qquad R\xi:= \sum_k
(R_k\xi) \otimes h_k\,,
\end{equation}
\begin{equation}\label{1.12}
K=H - \frac{{\rmi}}{2} \sum_k R_k^{\dagger }R_k\equiv H - \frac{{\rmi}}{2}
R^{\dagger }R\,.
\end{equation}
Then (see reference \cite{Partha92} Theorem 27.8 p.\ 228) there exists a unique
\emph{unitary} operator-valued adapted process $U(t)$ satisfying $U(0)
=\openone$ and
\begin{equation}\label{1.13} \fl
{\rmd} U(t) = \biggl\{ \sum_k R_k \,{\rmd} A_k^\dagger(t) + \sum_{kl}
\left(S_{kl}- \delta_{kl}\right) {\rmd} \Lambda_{kl}(t) - \sum_{kl}
R_k^{\dagger } S_{kl }\,{\rmd} A_l(t) -{\rmi} K\,{\rmd} t \biggr\}  U(t).
\end{equation}

The Bose fields introduced here represent a good approximation of
the electro\-magnetic field in the so called
\emph{quasi-monochromatic paraxial approximation}
\cite{Yuen,Bar90}. Now, ${\cal F}$ is interpreted as the Hilbert
space of the electromagnetic field; $A^\dagger_k(t)$ creates a
photon with state $h_k$ in the time interval $[0,t]$, $A_k(t)$
annihilates it, $\Lambda_{kk}(t)$ is the selfadjoint operator
representing the number of photons with state $h_k$ in the time
interval $[0,t]$. \emph{The operator $U(t)$ represents the
evolution operator} for a compound system (say an atom plus the
electromagnetic field), \emph{in the interaction picture with
respect to the free dynamics of the field.}

\subsubsection{The two-level atom.} The experiments \cite{Eze1}--\cite{Walther2}
involve or the hyperfine component of the $D_2$ line of sodium
with  up level $2P_{3/2}$, $F=3$ and down level $2S_{1/2}$, $F=2$,
or the levels $6s6p\ {}^1P_1$, $F=1$, and  $6s^2\ S_0$, $F=0$ of
${}^{138}$Ba; $F$ is the total angular momentum.

In order to describe an atom with two degenerate levels as in the experimental
situation, we take
\begin{equation}
\mathcal{ H}= \mathbb {C}^{2F_-+1} \oplus \mathbb{C}^{2F_++1} \,, \qquad
F_+=F_-+1\,,
\end{equation}
where $F_-$ is integer or semi-integer. We denote by $|F_\pm \,,\, M\rangle $,
$M=-F_\pm, \ldots , F_\pm$, the angular momentum basis in $\mathcal H$; the
parities of the states of the two levels must be opposed, let us say they are
$\epsilon_\pm $, with $\epsilon_+\epsilon_-=-1$. Let us denote by $P_\pm=
\sum_{M= -F_\pm}^{F_\pm} |F_\pm \,,\, M\rangle \langle F_\pm \,,\, M| $ the two
projection on the up or down states. Then the free energy $H$, contained in the
quantity \eref{1.12}, must be given by
\begin{equation}
H=\case 1 2 \,\omega_0 (P_+-P_-)\,, \qquad \omega_0>0\,,
\end{equation}
where the atomic frequency $\omega_0$ must already include the
Lamb shift.

\subsubsection{The photon space.}\label{photonspace}
In the approximations we are considering \cite{Yuen,Bar90}, the fields behave
as monodimensional waves, so that a change of position is equivalent to a
change of time and viceversa. Then, the space ${\cal Z}$ has to contain only
the degrees of freedom linked to the direction of propagation and to the
polarization. To describe a spin-1 0-mass particle we use the conventions of
Messiah \cite{Messiah} pp.\ 550, 1032--1037.

A c.o.n.s.\ in $\mathcal{Z}$ is given by $|j,m;\varpi \rangle \equiv \vec
\Theta_{jm}^\varpi$ , $j=1,2,\ldots$, $m=-j,-j+1,\ldots , j$, $\varpi=\pm 1$;
$jm$ is the total angular momentum, $\varpi=+1$ denotes electrical multipole,
$\varpi=-1$ denotes magnetic multipole and $(-1)^j \varpi$ is the parity. By
using the spherical harmonics $Y_{lm}(\theta,\phi)$ and the orbital angular
momentum operator $\vec \ell$, one has
\begin{equation}
\vec \Theta_{jm}^{-1}= \frac 1 {\sqrt{j(j+1)}} \, \vec \ell \, Y_{jm}\,, \qquad
\vec \Theta_{jm}^{+1}=\rmi \vec p\times  \vec  \Theta_{jm}^{-1}\,,
\end{equation}
where $\vec p$ is the direction versor given by $p_1=\sin \theta \, \cos \phi$,
$p_2= \sin \theta \, \sin \phi $, $p_3=\cos \theta$.

\subsubsection{The interaction.} Let us consider now the terms
with the creation and annihilation operators in the dynamical
equation \eref{1.13}; they must describe the absorption/emission
process. By asking spherical symmetry, parity conservation and
only electrical dipole contribution in $R$, we must have
\begin{equation}
R= \alpha \sum_{M=-F_+}^{F_+} |\epsilon_+; F_-;1;F_+,M\rangle
\langle F_+,M|, \qquad \alpha \in \mathbb{C}\,, \quad \alpha \neq
0\,,
\end{equation}
\begin{equation}\fl
|\epsilon_+; F_-;1;F_+,M\rangle = \sum_{m_1=-F_-}^{F_-}
\sum_{m_2=-1}^1 |F_-,m_1\rangle \otimes |1,m_2;+1\rangle \langle
F_-,m_1;1,m_2|F_+,M\rangle,
\end{equation}
where by $\langle F_-,m_1;1,m_2|F_+,M\rangle$ we denote the
Clebsch-Gordan coefficients (\cite{Messiah} pp.\ 560--563). The
interaction terms containing $R$ are responsible for the
spontaneous decay of the atom and, as we shall see, $|\alpha|^2$
turns out to be the natural line width. Note that
\begin{equation}
R_k=P_-R_k P_+\,, \qquad R^\dagger R = |\alpha|^2 P_+\,.
\end{equation}

The interaction term containing the $\Lambda$-process must give
the residual scattering when the atom is frozen in the up or down
level, so we take the unitary operator $S$ of the form
\begin{equation}\label{2.16}
S= P_+\otimes S^+ + P_-\otimes S^-\,, \qquad S^{\pm} \in
\mathcal{U}(\mathcal{Z})\,.
\end{equation}
Then, by spherical symmetry and parity conservation we must have
\begin{equation}
S^\pm = \sum_{j=1}^\infty \sum_{m=-j}^j \sum_{\varpi=\pm 1}
|j,m;\varpi\rangle \exp\{2 \rmi \delta_\pm(j;\varpi)\}\langle
j,m;\varpi|,
\end{equation}
$0\leq  \delta_\pm(j;\varpi) <2\pi $; the $\delta$'s are the phase
shifts for the atom frozen in the up or down level. Quantities
like $\omega_0$, $\alpha$, $\delta_\pm(j;\varpi)$ are
phenomenological parameters, or, better, they have to be computed
from some more fundamental theory, such as some approximation to
quantum electrodynamics.

\subsection{Some dynamical features of the model}
\label{master}

\subsubsection{The master equation.} Equation \eref{1.13} already contains the usual Markov
approximations; indeed, when the initial state of the field is a coherent
vector, then the atomic reduced dynamics turns out to be given by a master
equation without need of introducing further approximations
\cite{1HP84,Partha92,Bar90}. If $\rho_0$ is the initial atomic density matrix
on $\mathcal{H}$ and the coherent vector $e(f)$ is the state of the field, then
the atomic reduced statistical operator
\begin{equation} \label{2.2}
\rho(t;f) = \mathrm{Tr}_{{}_{\cal F}} \left\{ U(t) \bigl(\rho_0
\otimes |e(f) \rangle \langle e(f)|  \bigr)U(t)^\dagger  \right\}
\end{equation}
turns out to satisfy the master equation
\begin{equation}\label{2.5}
\frac{{\rmd}\ }{ {\rmd} t} \, \rho(t;f) = {\cal L}\big(f(t)\big)
[\rho(t;f)]\,,
\end{equation}
where the time dependent Liouvillian turns out to be
\numparts
\begin{eqnarray}
\fl {\cal L}\big(f(t)\big)[\rho] = -{\rmi} \left[H\big(f(t)\big),
\rho\right]
\nonumber \\
\lo{+} \frac 1 2 \sum_k \left( \left[ R_k \big(f(t)\big)\rho,
R_k\big(f(t)\big)^\dagger \right] +\left[ R_k\big(f(t)\big), \rho
R_k\big(f(t)\big)^\dagger \right]\right), \label{2.6a}
\end{eqnarray}
\begin{equation}\label{2.6b}
R_k\big(f(t)\big) = R_k +\langle h_k|S^+ f(t) \rangle P_+ +
\langle h_k|S^-f(t) \rangle P_- \,,
\end{equation}
\begin{equation} \label{2.6c}
H\big(f(t)\big) = H + \frac {\rmi} 2 \sum_{k} \Bigl( \langle S^-
f(t) |h_k \rangle  R_k  - \langle h_k | S^- f(t) \rangle
R_k^\dagger \Bigr).
\end{equation}
\endnumparts
Note that, when the field is in the Fock vacuum, only spontaneous
emission is present; indeed, for $f(t)=0$, the Liouvillian
\eref{2.6a} reduces to
\numparts
\begin{equation}\label{f=0}\fl
{\cal L}(0)[\rho] = - \frac{\rmi}2\, \omega_0 \left[P_+ -P_-\,,
\rho\right] - \frac 1 2 \left|\alpha\right|^2 \left( P_+ \rho +
\rho P_+ \right) + |\alpha|^2 \sum _{m=-1}^1 A_m \rho
A_m^\dagger\,,
\end{equation}
\begin{equation}\fl
A_m = \sum_{m_1=-F_-}^{F_-} |F_-, m_1 \rangle \langle
F_-,m_1;1,m|F_+,m_1+m\rangle \langle F_+,m_1+m|\,,
\end{equation}
\endnumparts
which describes decay according to the usual selection rules for
electric dipole and fixes the meaning of $|\alpha|^2$.

\subsubsection{The balance equation for the number of photons.}
Let us introduce the observables ``total number of photons in the time interval
$[0,t]$'' before and after the interaction with the atom
\begin{equation}\label{1.9}
N^{\rm in}(t)= \sum_k \Lambda_{kk}(t)\,, \qquad N^{\rm out}(t)=
U^\dagger(t) N^{\rm in}(t) U(t)\,.
\end{equation}
By using the computations on the increments of the ``output
fields'' given in \cite {Bar90}, we obtain in the present model
the balance equation
\begin{equation}\label{balance}
N^{\rm in}(t) +\frac 1 2 \,(P_+ - P_-)=N^{\rm out}(t)+ \frac 1 2
\,U(t)^\dagger (P_+ - P_-)U(t)\,.
\end{equation}
Such an equation says that the number of photons entering the system up to time
$t$ plus the photons stored in the atom at time 0 is equal to the number of
photons leaving the system up to time $t$ plus the photons stored in the atom
at time $t$. An equation for mean values similar to \eref{balance} has been
used  for restricting the possible forms of the interaction term in the simpler
model studied in \cite{BarL2000}.

\subsection{The laser and the equilibrium state}
\subsubsection{The phase diffusion model for the laser.} As in \cite{kim-man1,kim-man2}
we take as the state of the laser a mixture of coherent vectors
$\mathbb{E} \big[ |e(f) \rangle \langle e(f)| \big]$ with
\begin{equation}\label{2.19}\fl
f(t)= \exp \left\{-{\rmi}\left(\omega t + \sqrt{B} \, W(t)
\right)\right\}\theta(T-t)\, \lambda\,, \quad \lambda \in \mathcal {Z}\,, \quad
\omega>0\,, \quad B\geq 0\,,
\end{equation}
$W(t)$ is a standard Wiener process, $\theta$ is the usual step function and
$T$ is a large time; $T\to +\infty$ in the final results is always understood
and $\mathbb{E}$ means the classical expectation over the Wiener process. Note
that this model of the laser implies a Lorentz spectrum of bandwidth $B$;
indeed, for $T\to \infty$, one obtains
\begin{equation}
\frac {\hbar\omega} {2\pi} \int_{-\infty}^{+\infty} \rme^{\rmi \nu \tau }\,
\mathbb{E} [ \langle f(t) | f(t+\tau)\rangle ] \, \rmd \tau =
\hbar\omega\|\lambda\|^2\, \frac {B/(2\pi)} {(\nu-\omega)^2 + B^2/4}\,.
\end{equation}
The whole model is meaningful only for $\omega$ not too ``far'' from
$\omega_0$.

In order to describe a well collimated laser beam propagating
along the direction $z$ $(\theta=0)$ and with right circular
polarization, we take
\begin{equation}\label{3.30}
\lambda =  \alpha \Omega  \,{\rme}^{{\rmi} \delta} \lambda_+\,, \qquad
\Omega>0\,, \quad  \delta\in [0,2\pi)
\end{equation}
\begin{equation}
\vec \lambda_+ (\theta,\phi) = \frac{ 1_{[0,\Delta \theta]}(\theta)} {\Delta
\theta \sqrt{3\pi (1-\cos \Delta \theta)}} \left(- \frac 1 {\sqrt{2}}\right)
\left( \vec i + \rmi \vec j \right),
\end{equation}
where $1_{[0,\Delta \theta]}(\theta)=1$ for $0\leq \theta \leq \Delta \theta$,
$1_{[0,\Delta \theta]}(\theta)=0$ elsewhere; in all the physical quantities the
limit $\Delta \theta \downarrow 0$ will be taken. Note that the power of the
laser $\hbar \omega\|\lambda\|^2=\frac 2 3 \hbar\omega |\alpha|^2 \Omega^2/
{(\Delta\theta)^2}$ diverges for $\Delta \theta \downarrow 0$, because we need
a not vanishing atom-field interaction in the limit. In the following we shall
need the relation
\begin{equation}
\langle j,m;\varpi|\lambda_+\rangle = - \varpi \sqrt{ \frac {2j+1} {12} } \,
\delta_{m,1}
\end{equation}

The choice of a circular polarization is done in some experiments, because in
this way, in the long run, only two states are involved in the dynamics
(\cite{Walther1} p.\ 206) and the usual theory has been developed for
two-states systems \cite{Moll,kim-man1}; we shall see that also in our modified
model the right circular polarization pushes the atom in the states
$|F_-,F_-\rangle$, $|F_+,F_+\rangle$.

\subsubsection{The equilibrium state.} By a stochastic unitary transformation,
we obtain from \eref{2.2} and \eref{2.19} an atomic density matrix
\begin{eqnarray}\fl
\hat\rho(t) = \exp \left\{\frac{\rmi}2\left(\omega t + \sqrt{B} \, W(t)
\right)(P_+-P_-)\right\}\rho(t;f) \nonumber
\\ \lo\times
\exp \left\{-\frac{\rmi}2\left(\omega t + \sqrt{B} \, W(t)
\right)(P_+-P_-)\right\},\label{3.2}
\end{eqnarray}
which satisfies the ``Ito's'' master equation
\begin{equation}\label{randmaster}
\rmd \hat \rho(t) = \mathcal{L}_B\big[\hat \rho(t)\big]\rmd t + \frac \rmi 2\,
\sqrt{B} \bigl[ P_+-P_- , \hat \rho(t) \bigr]\rmd W(t)
\end{equation}
and whose mean value satisfies  the ordinary master equation
\begin{equation}\label{meanrho}
\rho(t)=\mathbb{E} \bigl[ \hat \rho(t) \bigr], \qquad \frac{{\rmd} \ }{{\rmd}
t}\, \rho(t) = \mathcal {L}_B \left[\rho(t)\right].
\end{equation}
The time independent Liouvillian entering equations \eref{randmaster} and
\eref{meanrho} is given by
\numparts
\begin{equation}\label{liouv}\fl
\mathcal{L}_B[\rho] = \mathcal{L}_0[\rho] + \frac B 8 \Bigl( \bigl[ \left(
P_+-P_- \right) \rho, P_+-P_- \bigr]+\bigl[ P_+-P_- , \rho \left(
P_+-P_-\right) \bigr]\Bigr) ,
\end{equation}
\begin{equation} \label{3.4a}\fl
{\cal L}_0[\rho] = -{\rmi} [\widehat H\,, \rho] +\frac 1 2 \sum_k \left(\left[
D(h_k)\rho\,,\, D(h_k)^{\dagger } \right]+ \left[ D(h_k)\,,\, \rho
D(h_k)^{\dagger } \right]\right),
\end{equation}
\begin{eqnarray} \fl
\widehat H = \frac 12\, (\omega_0 - \omega)(P_+ - P_-)  + \frac \rmi
2 \sum_k \left( \langle (S^- + \openone ) \lambda |h_k \rangle R_k - \langle
h_k | (S^- + \openone )\lambda \rangle R_k^\dagger \right) \nonumber
\\ \lo+ \frac \rmi 2 \left( \langle \lambda | S^+ \lambda \rangle - \langle S^+\lambda |
\lambda \rangle \right) P_+ +  \frac \rmi 2 \left( \langle \lambda | S^-
\lambda \rangle - \langle S^-\lambda | \lambda \rangle \right) P_-\,,
\label{3.4c}
\end{eqnarray}
\endnumparts
where $D(h)$ is a kind of \emph{effective dipole operator}
\begin{eqnarray}\nonumber
\lo{D(h)} &= \sum_k \langle h | h_k\rangle R_k + \langle h |\left(S^+ -
\openone\right)\lambda \rangle P_+ +\langle h |\left(S^- -\openone\right)
\lambda \rangle P_-
\\
&= \alpha \sum_{m=-1}^1 \langle h|1,m;+1\rangle A_m + \alpha \Omega \rme^{\rmi
\delta }\sum_{\epsilon = \pm } \langle h|\left( S^\epsilon -\openone \right)
\lambda_+ \rangle P_\epsilon \,. \label{3.4b}
\end{eqnarray}

Now, let us set
\numparts
\begin{equation}
P_\bot=\openone - |1,1;+1\rangle \langle 1,1;+1|\,, \qquad \varepsilon=
\mathrm{Im}\langle S^+\lambda_+|P_\bot S^-\lambda_+\rangle\,,
\end{equation}
\begin{equation}\label{ggg}
 g_\pm = P_\bot \big(S^\pm - \openone\big) \lambda_+ \,, \qquad \Delta g
= g_+-g_-\,,
\end{equation}
\begin{equation}\label{delta}
\delta_\pm = \delta_\pm(1;+1)\,, \qquad s=\delta_+ - \delta_-\,,
\end{equation}
\endnumparts
We have to ask $|\varepsilon|< +\infty$, $\|g_\pm\|<+\infty$,  for
$\Delta \theta \downarrow 0$; roughly speaking, $S_\pm$ must
introduce small corrections even when the norm of $\lambda_+$
diverges. Then, we have \numparts
\begin{eqnarray}\fl\label{elba}
P_+ &\mathcal{L}_B[\rho] P_+ = -|\alpha|^2 P_+\rho P_+ -L^{-\dagger}P_- \rho
P_+ - P_+ \rho P_- L^-\,,
\\ \fl
P_- &\mathcal{L}_B[\rho] P_- = |\alpha|^2\sum_{m=-1}^1 A_m P_+ \rho P_+
A_m^\dagger + L^- P_+ \rho P_- + P_- \rho P_+ L^{-\dagger}\,,
\\ \nonumber \fl
P_-&\mathcal{L}_B[\rho]P_+ = \left( P_+
\mathcal{L}_B\left[\rho^\dagger\right]P_-\right)^\dagger = L^+ P_+ \rho P_+ -
P_- \rho P_- L^-
\\ \fl
{}&+ \left[ \rmi \left( \omega_0 - \omega + \mathrm{Im} \langle
S^+ \lambda |S^- \lambda \rangle\right) - \case 1 2 \left(
|\alpha|^2 +B + \|(S^+ - S^-) \lambda \|^2 \right) \right] P_-\rho
P_+, \label{elbc}
\end{eqnarray}
\endnumparts

\numparts
\begin{eqnarray}
L^\pm = \sum_k \langle S^\pm \lambda| h_k\rangle R_k = - \frac 1 2 \,|\alpha|^2
\Omega \,\rme^{ - \rmi (\delta + 2 \delta_\pm )} A_1\,,
\\
\big(S^\pm - \openone\big) \lambda_+  = -\rmi \rme^{\rmi \delta_\pm }\sin
\delta_\pm |1,1;+1\rangle + g_\pm\,,
\\
\mathrm{Im} \langle S^+ \lambda | S^- \lambda \rangle =|\alpha|^2 \Omega^2
\left(\varepsilon - \case 1 4 \sin 2s\right),
\\
\left\| \left(  S^+ - S^-\right) \lambda \right\|^2 =|\alpha|^2 \Omega^2 \left(
\sin^2 s + \|\Delta g \|^2 \right),
\end{eqnarray}
\endnumparts

Then, one can check that the equilibrium state $\rho_\infty$ for
the master equation \eref{meanrho} is supported by the two extreme
states
\begin{equation}
|1\rangle= |F_+,F_+ \rangle \,, \qquad |2\rangle = |F_-,F_- \rangle.
\end{equation}
By using these two states the density matrix  $\rho_\infty$ is given by
\begin{equation}\label{rho}
\rho_\infty = \left( \begin{array}{cc}  \Omega^2 d_1  & \Omega \exp\left[
\rmi \left( \delta +2 \delta_-\right)\right] d_2
\\ &  \\ \Omega \exp\left[ -\rmi \left( \delta +2
\delta_-\right)\right] d_3 & 1-\Omega^2 d_1
\end{array} \right),
\end{equation}
\begin{equation}\label{3.1b}
\bi{d}= \left( \begin{array}{c} d_1\\ d_2 \\  d_3
\end{array} \right)\,,
\qquad \bi{G} \bi{d} = \bi{w}\,,\qquad \bi{w}=\frac{ 1} 2 \left(
\begin{array}{c} 0\\ 1  \\ 1
\end{array}\right) ,
\end{equation}
\begin{equation}\label{3.7c}
\bi{G}= \left( \begin{array}{ccc} 1 & -1/2 & -1 /2
\\ & & \\
\Omega^2 \rme^{\rmi s} \cos s & b & 0
\\ & & \\
\Omega^2 \rme^{-\rmi s} \cos s  & 0 &  \overline b \end{array} \right),
\end{equation}
\begin{equation}
b= \case 1 2 \left[ 1+y + \Omega^2 \left( \|\Delta g \|^2 + \sin^2 s
\right)\right] - \rmi \left( \widetilde  z + \case 1 4 \Omega^2 \sin 2s
\right),
\end{equation}
\begin{equation}\label{detuning}
\widetilde  z = z - \Omega^2 \varepsilon\,, \qquad z = \left( \omega - \omega_0
\right)/ |\alpha|^2\,, \qquad y = B/ |\alpha|^2\,.
\end{equation}
It turns out
\begin{equation}
d_1=\mathrm{Re}\,d\,, \qquad d_2 =d\,, \qquad d_3 = \overline d\,,
\end{equation}
\begin{equation}\label{stst}
d= \frac { 1+y +\Omega^2 \left(\|\Delta g\|^2 + \sin^2 s \right) + \rmi \left(
2\widetilde  z - \Omega^2 \sin s\, \cos s \right)} { 4 \widetilde z ^2 +
\Gamma^2 }\,,
\end{equation}
\begin{equation}\fl\label{Gamma2}
\Gamma^2 = \left( 1+ y+\Omega^2 \left\|  \Delta g \right\|^2 \right)^2
+\Omega^2 \left( 2+ 2y + 2\Omega^2 \left\|  \Delta g \right\|^2 +\Omega^2 \sin
^2 s \right). \label{3.36m}
\end{equation}

\section{The fluorescence spectrum}\label{flspectrum}
\subsection{Heterodyne detection}\label{heter}

\subsubsection{The output current.} A way to obtain the spectrum of the stimulated
atom is by means of \emph{balanced heterodyne detection}; in such a scheme the
light emitted by the atom is made to interfere with the light of a strong
monochromatic laser (the local oscillator) and then detected by a couple of
photocounters \cite{SYMM}. A mathematical description of such a detection
scheme has been given inside the theory of continuous measurements
\cite{Bar90,Goslar}; the output current of the detector is represented by the
operator
\begin{equation}\label{5.1}\fl
I(\nu,h;t) = \int_0^t F(t-s) {\rme}^{{\rmi} \nu s} \, {\rmd} A(h;s) +
\mathrm{h.c.}\,, \qquad {\rmd} A(h;t) = \sum_k \langle h|h_k\rangle
{\rmd}A_k(t)\,,
\end{equation}
where $F(t)$ is the detector response function, say
\begin{equation}\label{5.2}
F(t) = k_1 \sqrt{\frac{\widetilde \gamma}{4\pi}}\, \exp\left( -\frac{\widetilde
\gamma}{2}\, t \right), \qquad \widetilde \gamma>0\,,
\end{equation}
$k_1\neq 0$ has the dimensions of a current, $\nu$ is the frequency of the
local oscillator and $h\in \mathcal Z$, $\|h\|=1$. Any information on the
localization and on the polarization of the detector is contained in $h$. We
assume that the detector spans a small solid angle, so that $h$ is given by
\begin{equation}\label{5.4}
h(\theta^\prime, \phi^\prime) = \frac{ 1_{\Delta \Upsilon}
(\theta^\prime,\phi^\prime)}{\sqrt{|\Delta \Upsilon|}}\, \vec e(\theta^\prime,
\phi^\prime) ,
\end{equation}
where $\Delta \Upsilon$ is a small solid angle around the direction
$(\theta,\phi)$, $\Delta \Upsilon \downarrow \{(\theta,\phi)\}$, $| \Delta
\Upsilon | \simeq \sin \theta \, {\rmd }\theta \, {\rmd} \phi$ and $\vec e$ is
a complex polarization vector, $\left|\vec e(\theta^\prime, \phi^\prime)
\right|=1$. Moreover, we assume that the transmitted wave does not reach the
detector, i.e.\ $\theta>0$ and so
\begin{equation}\label{5.22}
\langle h | \lambda\rangle =0\,.
\end{equation}

From the canonical commutation relations for the fields one has that $I(\nu, h;
t_1)$ and $I(\nu, h; t_2)$ are compatible observables for any choice of the
times; the same holds in the Heisenberg picture. Then, a theory of continuous
observation of $I(\nu,h;t)$, $t\geq 0$, can be developed \cite{Bar90,BPag} and
a probability law for the stochastic process ``output current'' can be
constructed; however, to obtain the spectrum we do not need the full theory,
but only the second moment of $I$. Here and in the following for the
expectation of any operator $M$ we shall use the notation
\begin{equation}\label{5.11}
\langle \langle M\rangle \rangle = \mathbb{E} \left[
\mathrm{Tr}_{{}_{\mathcal{H}\otimes \mathcal{F}}}\left\{M \left(\rho_0 \otimes
|e(f)\rangle \langle e(f) | \right) \right\}\right].
\end{equation}

\subsubsection{The power spectrum.} In the long run the mean of the output
power is given by
\begin{equation}\label{5.5}
P(\nu,h) = \lim_{T \to +\infty} \frac{k_2}{T} \int_0^T \big\langle\big\langle
U(T)^\dagger \big( I(\nu,h;t)\big)^2 U(T)\big\rangle\big\rangle \,{\rmd}t\,;
\end{equation}
$k_2>0$ has the dimensions of a resistance, it is independent of $\nu$, but it
can depend on the other features of the detection apparatus. As a function of
$\nu$, $P(\nu,h)$ gives the \emph{power spectrum} observed in the ``channel
$h$''; in the case of the choice (\ref{5.4}) it is the spectrum observed around
the direction $(\theta,\phi)$ and with polarization $\vec e$.

The mean power $P(\nu,h)$, under condition \eref{5.22}, can be expressed as
\begin{equation}\label{5.8}
P(\nu,h) = \frac {k_1^2k_2}{4\pi} + k_1^2k_2\Sigma(x;h)\,,
\end{equation}
\begin{equation}\label{5.16}\fl
\Sigma(x;h) = \frac {\Omega^2} {2\pi} \int_0^{+\infty} {\rmd}t\,
\rme^{ -q t}\, \mathrm{Tr}_{{}_{\mathcal{H}}} \left\{ \widetilde
D(h)^\dagger \,{\rme}^{ \mathcal{ K} t } \left[ \widetilde D(h)
\rho_{\infty} \right]\right\} + \mathrm{c.c.},
\end{equation}
\begin{equation}\label{6.5}
x= \frac{ \nu-\omega_0} {|\alpha|^2}\,,  \qquad  \gamma = \frac
{\widetilde \gamma} {|\alpha|^2}\,, \qquad q= \frac{\gamma +y} 2 +
\rmi (x-z)\,.
\end{equation}
\begin{equation}
\widetilde D(h) = \rme^{-\rmi \delta}D(h)/(\alpha \Omega)\,,
\end{equation}
\begin{equation}
|\alpha|^2 \mathcal{K}[\rho] = \mathcal{L}_B[\rho] - B P_+\rho P_-
+ B P_- \rho P_+\,;
\end{equation}
$D(h)$ is given by equation (\ref{3.4b}), $\rho_\infty$ by
\eref{rho}, $\mathcal{L}_B$ by \eref{elba}--\eref{elbc}, $y$ and
$z$ by \eref{detuning}.

In the expression \eref{5.8} of the power the term $k_1^2k_2/(4\pi)$ appears,
independent of $\nu$; it is a white noise contribution due to the detection
scheme, coming out mathematically from the canonical commutation relations for
the fields and known as ``shot noise''. The semigroup $\exp[\mathcal{K}t]$ is
trace preserving, but not positivity preserving, while $\rho_\infty$ is the
equilibrium state of $\mathcal{L}_B$, which is a bona-fide Liouvillian, because
it can be written in the Lindblad form \eref{liouv}, \eref{3.4a}.

\subsubsection{Sketch of the proof of equations \eref{5.8} and \eref{5.16}.}
The computations are similar to those of reference \cite{BarL2000}, but now not
only QSC but also classical stochastic calculus plays a role. Here we give only
the main intermediate steps; a more detailed account can be found in
\cite{Pero}, where however the degeneracy of the levels is not considered.

The first step is to rearrange in normal order the field operators
involved in \eref{5.5}; in this way the constant shot noise
appears and the spectrum reduces to
\numparts
\begin{eqnarray}\fl
\Sigma(x;h)&= \lim_{T\to +\infty} \frac{1}{ 2\pi T }
\Big\langle\Big\langle \int_0^T {\rmd} A^{\dagger}_{\mathrm{out}}
(h;t) \int_0^t {\rmd} A_{\rm out}(h;s) \, {\rme}^{ - \left( \frac
{\widetilde \gamma} 2 +{\rmi} \nu \right) (t-s) }
\Big\rangle\Big\rangle + \mathrm{c.c.}
\\ \fl
&= \lim_{T\to +\infty} \frac{1}{ 2\pi T } \Big\langle\Big\langle
\int_0^T {\rmd} A^{\dagger}_{\mathrm{out}}(h;t)\rme^{-\rmi \nu t}
\int_0^T {\rmd} A_{\rm out}(h;s) \rme^{\rmi \nu s} \, {\rme}^{ -
\frac {\widetilde \gamma} 2 |t-s| } \Big\rangle\Big\rangle \geq 0;
\end{eqnarray}
\endnumparts
the second expression shows the positivity of $\Sigma(x;h)$, while
the first one is useful for the successive developments. The
output fields \cite{Bar90} are given by $A_{\rm out}(h;t) =
U(t)^\dagger A(h;t)U(t)$ and we have
\begin{equation}\label{Aout}
\rmd A_{\rm out}(h;t)=  \sum_k \langle h|h_k\rangle U(t)^\dagger
\biggl(\sum_l S_{kl} \rmd A_l(t) + R_k \rmd t\biggr) U(t)\,.
\end{equation}

The second step is to show, by using \eref{Aout}, \eref{3.2},
\eref{3.4b}, \eref{5.22} and the factorization properties of Fock
space, that
\begin{eqnarray}\fl \nonumber
\Sigma(x;h)= \lim_{T\to +\infty } \frac 1 {2\pi T } \int_0^T \rmd
t \int_0^t \rmd s \,\exp\left\{ - \left[ \frac {\widetilde
\gamma}2 +\rmi (\nu-\omega)\right] (t-s)\right\}
\\ \lo\times \mathbb{E}\Big[ \exp\left\{
\rmi \sqrt{B} \bigl[ W(t) - W(s) \bigr] \right\}
\mathrm{Tr}_{{}_{\mathcal{H}\otimes\mathcal{F}}} \Bigl\{ D(h)^\dagger C(t,s)
D(h) \nonumber
\\ {}\times \left( \hat \rho(s) \otimes |e(f)\rangle \langle e(f)| \right)
C(t,s)^\dagger \Bigr\}\Bigr] + \mathrm{c.c.}\,,
\end{eqnarray}
\begin{eqnarray} \fl \nonumber
C(t,s) = \exp\left\{ \frac \rmi 2 \left[ \omega t + \sqrt{B} \, W(t)
\right]\left(P_+ - P_-\right)\right\} U(t)U(s)^\dagger
\\ \lo\times \exp\left\{- \frac
\rmi 2 \left[ \omega s + \sqrt{B} \, W(s) \right]\left(P_+ -
P_-\right)\right\}.
\end{eqnarray}

Finally an interplay between classical stochastic calculus and the quantum
regression theorem, which holds for a dynamic like $C(t,s)$, gives the
expression \eref{5.16}.

\subsubsection{The angular distribution of the spectrum.}
Let us stress that $\rho_\infty $ is supported by the two states $|1\rangle=
|F_+,F_+ \rangle $, $|2\rangle = |F_-,F_- \rangle$ and that all the operations
involved in \eref{5.16} leave the span of $|1\rangle$, $|2\rangle$ invariant.
So we are left with computations involving $2\times 2$ matrices; from now on we
suppress the index $\mathcal{H}$ in the trace and we use the Pauli matrices
$\sigma_\pm$, $\sigma_z$, \ldots\ In particular we have $P_\pm = \frac 1 2
\left( 1 \pm \sigma_z\right) $ and
\begin{equation}
\widetilde D(h)= \langle h|1,1;+1\rangle \, D_1+ \langle h| g_+\rangle P_+ +
\langle h| g_-\rangle P_-\,,
\end{equation}
\begin{equation}
D_1= \frac 1 \Omega \, \rme^{-\rmi \delta}\, \sigma_- - \rmi \rme ^{\rmi
\delta_+} \sin \delta_+ \,P_+ - \rmi \rme^{\rmi \delta_-}\sin \delta_- \,P_-\,.
\end{equation}

To obtain the angular dependence of the spectrum, let us introduce
the states $h_\pm$ concentrated around $(\theta,\phi)$ and with
right/left circular polarization, given by equation \eref{5.4}
with $\vec e=\vec e_\pm $, where $\vec e_\pm(\theta,\phi)^\dagger
= \frac {\exp(-\rmi \phi)}  {\sqrt{2}} \bigl( -\rmi \sin \phi \mp
\cos \theta \,\cos \phi, \, \rmi \cos \phi \mp \cos \theta \,\sin
\phi, \,\pm \sin \theta \bigr) $. Then, the two angular spectra
$\Sigma_\pm(x; \theta)$, which, by the cylindrical symmetry of the
problem, do not depend on $\phi$, are given by
\begin{equation}\fl
\Sigma_\pm (x;\theta) = \frac 1{|\Delta \Upsilon|}\, \Sigma
(x;h_\pm)=\frac {\Omega^2}  {2\pi} \int_0^{+\infty} {\rmd}t\,
\rme^{ -q t}\, \mathrm{Tr} \left\{ D_\pm(\theta)^\dagger
\,{\rme}^{ \mathcal{ K} t } \left[  D_\pm(\theta) \rho_{\infty}
\right]\right\} + \mathrm{c.c.},
\end{equation}
\begin{equation}
 D_\pm(\theta)= \pm \frac 1 4 \sqrt {\frac 3 {2\pi}} \left( 1 \pm \cos \theta
 \right) D_1 + \sum_{\epsilon=\pm}g_\epsilon (\theta ; \pm )P_\epsilon\,.
\end{equation}
The functions $g_\epsilon (\theta ; \pm )= \langle h_\pm
|g_\epsilon\rangle \big/ \sqrt{|\Delta \Upsilon|}$ depend on
$S^\epsilon$, but after all they are free parameters of the
theory: they are square integrable $\theta$-functions, satisfying
the constraint $\int_0^\pi \sin \theta \left[ \left( 1+ \cos
\theta \right) g_\epsilon (\theta;+) - \left( 1- \cos \theta
\right) g_\epsilon (\theta;-)\right]\rmd \theta=0$, coming from
the orthogonality of $g_\epsilon$ to $|1,1;+1\rangle$, see
\eref{ggg}.

Then, by integrating over the whole solid angle, one gets the
total spectrum
\begin{equation}\label{totalS}\fl
\Sigma(x)= \int_0^\pi \rmd \theta \sin \theta \int_0^{2\pi} \rmd
\phi \Bigl(\Sigma_+(x; \theta)+\Sigma_-(x; \theta)\Bigr)= \sum_k
\Sigma(\nu;h_k)\,,
\end{equation}
where $\{h_k\}$ is any c.o.n.s.\ in $\mathcal{Z}$. When one has
$g_\epsilon=0$, as in the usual case, one gets
\begin{equation}
\Sigma_\pm(x; \theta)= \frac 3 {8\pi } \left( \frac{1\pm \cos
\theta} 2 \right)^2 \Sigma(x)\,.
\end{equation}
For $g_\epsilon\neq 0$ the $x$ and $\theta$ dependencies do not
factorize. In the experiments one measures something proportional
to $\Sigma_+(x; \theta)+\Sigma_-(x; \theta)$ for $\theta$ around
$\pi/2$; this quantity fails to be proportional to $\Sigma(x)$
only by the presence of some terms which we expect to be small and
which are not qualitatively different from the other terms in
$\Sigma(x)$. So, for simplicity, we shall study only the total
spectrum \eref{totalS}.

\subsection{The total spectrum}
By choosing in \eref{totalS} a basis with $h_1= |1,1;+1\rangle$, $h_2= \Delta g
/ \|\Delta g\|$, $h_3= \|\Delta g\|^{-1} \left[ \|g_-\|^2 \|\Delta g\|^2 -
|\langle \Delta g |g_-\rangle |^2\right]^{-1/2} \left [  \|\Delta g\|^2 g_- -
\langle \Delta g |g_-\rangle \Delta g \right]$, we get
\begin{eqnarray}\fl\nonumber
\sum_k&\mathrm{Tr}\left\{ \widetilde D(h_k)^\dagger
\rme^{\mathcal{K}t}\left[ \widetilde D(h_k) \rho_\infty\right]
\right\}= \mathrm{Tr}\left\{  D_1^\dagger
\rme^{\mathcal{K}t}\left[ D_1 \rho_\infty\right] \right\}+ \langle
g_-|\Delta g \rangle \,\mathrm{Tr}\left\{ P_+ \rho_\infty \right\}
\\ \fl {}&+\langle\Delta g|g_- \rangle
\,\mathrm{Tr}\left\{ P_+ \rme^{\mathcal{K}t}\left[
\rho_\infty\right] \right\}+\|g_-\|^2 + \|\Delta g\|^2
\,\mathrm{Tr}\left\{  P_+ \rme^{\mathcal{K}t}\left[
P_+\rho_\infty\right] \right\} . \label{traceD}
\end{eqnarray}
Then, from \eref{5.16}, \eref{totalS}, \eref{traceD}, we get the
final expression of the \emph{total spectrum}
\begin{eqnarray}\fl\nonumber
\Sigma(x) = \frac{\Omega^2}{2\pi} &\bigg[\frac 1 { q} \left(v_3 +
\rmi \rme^{\rmi \delta_-} \sin \delta_- + \rmi \Omega^2 v_1
\rme^{-\rmi s} \sin s \right) \bigl(d_2 - \rmi \rme^{-\rmi
\delta_-} \sin \delta_-
\\ \fl \nonumber
&- \rmi \Omega^2 d_1 \rme^{\rmi s} \sin s \bigr) +  \rme^{-\rmi
\delta_-} \sin \delta_+ \left (\Omega^2 c_1 \sin s -\rmi
\rme^{\rmi s} c_3\right)
\\ \fl \nonumber &+\left(d_1 + \rmi d_3
\rme^{\rmi s}\sin s \right) \left (u_3 + \rmi \Omega^2 u_1
\rme^{-\rmi s} \sin s\right) + \frac {1} { q} \left\langle g_-
\big| g_-+ \Omega^2 d_1 \Delta g \right \rangle
\\ \fl &+
\Omega^2 \left\langle \Delta g |g_+ \right\rangle c_1 -\Omega^2
d_3 \|\Delta g \|^2 u_1 +\frac {\Omega^2} { q} \left\langle \Delta
g \big| g_-+ \Omega^2 d_1 \Delta g \right \rangle v_1 +
\mathrm{c.c.}\bigg],\label{spectrum}
\end{eqnarray}
\begin{equation}\fl\label{vcu}
\bi{v}= \frac 1{\bi{K}+q}\,\bi{w}\,, \qquad \bi{c} = \frac
1{\bi{K}+q} \, \bi{d} \,, \qquad \bi{u}= \frac 1{\bi{K}+q}\left(
\begin{array}{c} 0 \\ 0 \\ 1\end{array} \right),
\end{equation}
\begin{equation}\label{K}
\bi{K}= \bi{G}+ y \left( \begin{array}{ccc} 0 & 0 & 0
\\ 0& 1 & 0 \\ 0 & 0 & -1
\end{array} \right).
\end{equation}
By integrating over the reduced frequency $x$, we get the strength
of the spectrum
\begin{eqnarray}\fl\nonumber
\int_{-\infty}^{+\infty} \Sigma(x) \,\rmd x = \Omega^2 \Bigl\{d_1 \left(1+
\Omega^2 \sin^2 \delta_+\right)+ \left(1-\Omega^2 d_1 \right) \left(\sin^2
\delta_-+\|g_-\|^2\right)
\\  + \mathrm{Re}\left[ d_2\left(\rme^{2\rmi \delta_-}-1\right) \right]  + \Omega^2d_1  \| g_+\|^2\Bigr\}.
\end{eqnarray}

Let us recall that $|\alpha|^2$ is the natural line width, $\Omega^2$ is
proportional to the laser intensity, $z=\left( \omega - \omega_0 \right)\big /
|\alpha|^2$ is the reduced detuning, to which also the parameter $\widetilde z=
z - \Omega^2 \varepsilon$  is linked, $y= B/|\alpha|^2$ is the reduced laser
bandwidth \eref{detuning}, $\Omega^2 |\alpha|^2 \varepsilon$ is an intensity
dependent shift, $x= \left( \nu-\omega_0\right)/ |\alpha|^2$ and $ \gamma=
\widetilde \gamma /|\alpha|^2$ are the reduced frequency and the reduced
instrumental width, respectively, and $q= \rmi (x-z) + \left(\gamma+y\right)/2$
\eref{6.5}. The other quantities entering all these equations are given by the
equations \eref{ggg}, \eref{delta}, \eref{3.1b}--\eref{Gamma2}; in particular,
$s=\delta_+-\delta_-$ and $\Delta g = g_+-g_-$. Let us note that $\varepsilon$,
$\delta_\pm$, $\left\|  g_\pm \right\|^2$, $\langle g_+|g_-\rangle$ are
parameters linked to the $S^\pm$ scattering matrices, satisfying the
constraints ``$\big|\langle g_+ |g_-\rangle \big| \leq \left\|g_+\right\|
\left\| g_-\right\|$'' and ``$\|\Delta g\|=0 \ \Rightarrow \ \varepsilon=0$''
(apart from this relation $\varepsilon$ is an independent parameter of the
model). One can check that the spectrum $\Sigma(x)$ is invariant under the
transformation: $x\to - x$, $z\to - z$, $\varepsilon\to - \varepsilon$,
$\delta_\pm \to -\delta_\pm$, $\langle g_-|g_+\rangle \to \langle
g_+|g_-\rangle$.

\subsubsection{The case of a monochromatic laser, $y=0$.} When the laser
bandwidth vanishes the spectrum \eref{spectrum} reduces to the one
obtained in \cite{BarL2000}, where the polarization of light and
the level degeneracy were not considered; if also $S^\pm=\openone$
is taken, it reduces to the Mollow spectrum \cite{Moll}. A
particularity of the $y=0$ case is that the spectrum can be
decomposed into an elastic and an inelastic part, which in turn
are proportional to the elastic and inelastic electromagnetic
cross sections, as explained in \cite{BarL2000}; for $y>0$ such a
decomposition has no meaning. Figures 2 and 3 of \cite{BarL2000}
compare the spectra (for $\gamma = 0.6$ and for various detunings
and laser intensities) in the Mollow case and in the modified case
with $\delta_+=-0.03$, $\delta_-=0.13$, $\|g_\pm \|^2 = 0.005$,
$\langle g_+| g_- \rangle = - 0.005$, $\varepsilon=-0.001$; the
two figures have $x-z$ in the horizontal axis and
$\Sigma/\Omega^2$ in the vertical one. The parameters have been
chosen in such a way that the modified spectrum would not too
different from the Mollow one in the resonant case.  In plotting
the spectrum a factor $1/2$ has been inserted by error in the
inelastic part, so the two figures have only a qualitative
meaning; in any case, one sees the symmetry of the Mollow spectrum
which is lost in the modified case.

\begin{figure}
\begin{center}
\includegraphics*{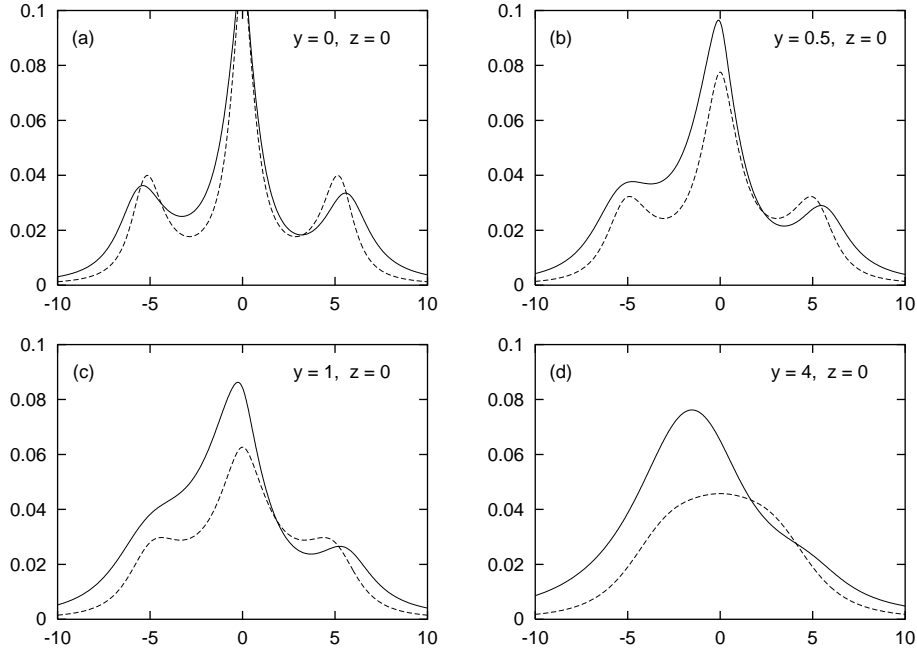}
\end{center}
\caption{The spectrum $\Sigma(x)$ for $\gamma=0.6$, $\Omega^2 = 28$, $z=0$ and
laser bandwidths $y=0,\ 0.5,\ 1, \ 4$. Solid lines: modified model with the
parameters given in section \ref{parameters}; dashed lines: usual model.}
\label{fig1}
\end{figure}

\subsubsection{The case $S^\pm =\openone$.} Let us recall that the usual model, with only
the absorption/emission process, corresponds to $\delta_\pm=0$,
$g_\pm = 0$, $\varepsilon=0$, $z=\widetilde z$, $s=0$. In this
case we obtain
\begin{equation} \label{5.49a}
\Sigma(x) = \frac {\Omega^2}{2\pi}\left[ \frac 1 q \, v_3 d+ \left(2 v_3  +
\frac {4\Omega^2} N \right)\mathrm{Re}\,d \right]+\mathrm{c.c.}\,,
\end{equation}
\begin{equation}\label{5.49b}
d= \frac {1+y +2\rmi z} {4z^2 +\Gamma^2}\,, \qquad \Gamma^2 =
(1+y)(1+y+2\Omega^2)\,,
\end{equation}
\begin{equation}
v_3= \left[ 2+\gamma +y +2\rmi \left(x-z\right)\right] \left[
1+\gamma +4y +2\rmi \left(x-2z\right)\right]\big/ N\,,
\end{equation}
\begin{eqnarray} \nonumber
N &= 4\Omega^2\left[ 1+\gamma +2y+ 2 \rmi(x-z) \right] +\left[
2+\gamma +y +2\rmi \left(x-z\right)\right]
\\ &\times\left[ 1+\gamma +4y
+2\rmi \left(x-2z\right)\right] \left( 1+\gamma + 2 \rmi x
\right).
\end{eqnarray}
Now, the spectrum $\Sigma(x)$ is invariant under the
transformation: $x\to - x$, $z\to - z$.

If we put also $\gamma=0$, which means that the instrumental width
is negligible, then one can check that the fluorescence spectrum
$\Sigma(x)$ coincides exactly (apart from the different
normalization) with the spectrum computed by Kimble and Mandel
\cite{kim-man1,kim-man2}; when also the laser bandwidth $y$
vanishes, the equations above reduce once again to the Mollow
spectrum \cite{Moll}. One of the more interesting results of
Kimble and Mandel is that a not vanishing laser bandwidth renders
asymmetric the Mollow spectrum and enhances the side peak near
$\omega_0$, as shown in the figures of \cite{kim-man1}. In our
notations, the figures 2, 3, 4 of \cite{kim-man1} give, for
$\gamma=0$, $\pi \Sigma$ on the z-axis versus $2(x-z)$ on the
x-axis and $y$ on the y-axis; figure 2 is for $\Omega^2=0.0025$
and $z=-2.5$, figure 3 for $\Omega^2=25$ and $z=0$, figure 4 for
$\Omega^2=25$ and $z=-1.5$. The figures 5 and 6 of \cite{kim-man1}
give, for $\gamma=0$, $\pi \Sigma$ on the z-axis versus $2(x-z)$
on the x-axis and $2z$ on the y-axis; figure 5 is for
$\Omega^2=25$, $y=0.5$ and figure 6 is for $\Omega^2=2500$,
$y=10$.

\subsubsection{Low intensity laser.} For a laser of low intensity the
spectrum reduces to
\begin{eqnarray}\fl\nonumber
\left. \frac {\pi\Sigma(x)}{2\Omega^2}\right|_{\Omega^2=0}= (\gamma+y) \left|
c(x-z, \gamma +y)\right|^2  \left( \|g_-\|^2 +  \frac{\gamma } {\gamma+y}
\left|a(y,z)\right|^2  \right)
\\ \nonumber \lo{+}
y \big| c(x-z,\gamma +y) a(y,z) + c(-x, 1+\gamma) c(-z,1+y)\big|^2
\\ \lo{+}
\gamma y \left| c(-x,1+\gamma)\right|^2 \left| c(-z, 1 +y)\right|^2 ,
\label{omega=0}
\end{eqnarray}
\begin{equation}\label{c()}\fl
a(y,z) = c(-z,1+y) -\rmi \rme^{-\rmi \delta_-} \sin \delta_- \,, \qquad
c(x,\Delta) = \frac 1 {\Delta +2\rmi x}\,.
\end{equation}
In this case one sees well why a decomposition in elastic and
inelastic parts is not possible for $y>0$. Indeed, one has two
peaks, one centered at $\nu=\omega$ (laser frequency) and one
centered at $\nu=\omega_0$ (atomic frequency); the first peak can
be interpreted as coming from an elastic channel and the second
one as coming from an inelastic channel, but the second term in
the r.h.s.\ of \eref{omega=0} gives rise to an interference term
between the two channels.

\begin{figure}
\begin{center}
\includegraphics*{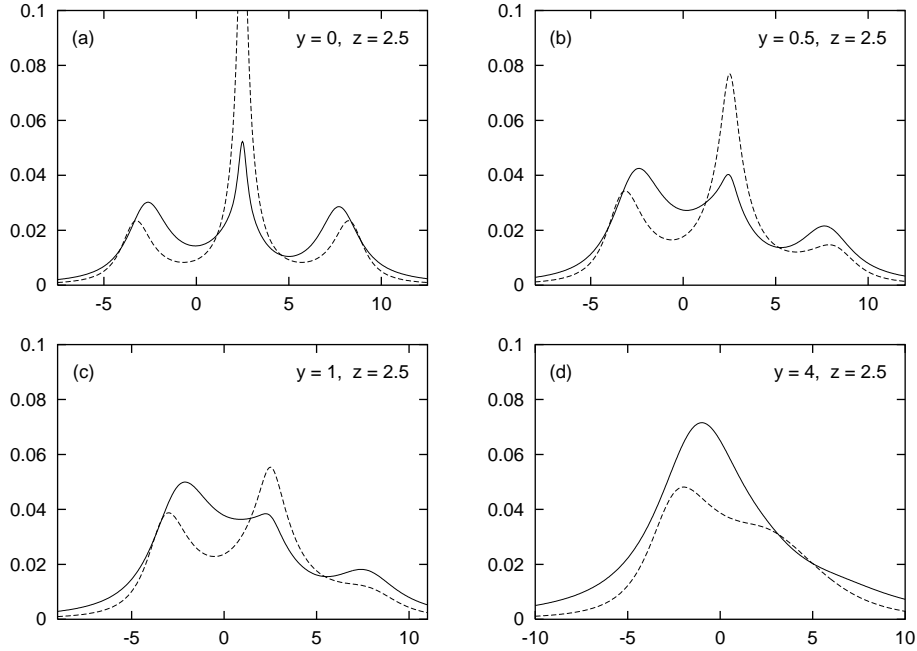}
\end{center}
\caption{The spectrum $\Sigma(x)$ for $\gamma=0.6$, $\Omega^2 = 28$, $z=2.5$
and  laser bandwidths $y=0,\ 0.5,\ 1, \ 4$. Solid lines: modified model with
the parameters given in section \ref{parameters}; dashed lines: usual model.}
\label{fig2}
\end{figure}

\subsubsection{The broadband case.} \label{broadband} For a very
large bandwidth, $y\gg x,\ z, \ \Omega^2$, the spectrum becomes
independent from the detuning and reduces to
\begin{equation}\fl
\lim_{y\to \infty} \frac {\pi y}{2\Omega^2}\, \Sigma(x) = \left|
c(x-\eta ,1+\kappa)+\rmi \rme^{\rmi \delta_-}\sin \delta_-
\right|^2 + \kappa \left|c(x-\eta ,1+\kappa)\right|^2 +
\|g_-\|^2\,,
\end{equation}
where $c(x,\Delta)$ is given in \eref{c()} and
\begin{equation}
\eta = \Omega^2\left( \varepsilon - \frac 1 4 \, \sin 2 s\right),
\qquad \kappa = \gamma + \Omega^2 \left( \sin^2 s + \|\Delta g\|^2
\right).
\end{equation}
Note the $\Omega$-dependent position of the peak, which can be
interpreted as a light shift. In the \emph{usual case} $\Sigma(x)$
becomes purely Lorentzian, centered in zero and with
$\Omega$-independent width:
\begin{equation}
\lim_{y\to \infty} \frac {\pi y}{2\Omega^2}\, \Sigma(x) = \frac
{1+\gamma}{(1+\gamma)^2 +4x^2} \,.
\end{equation}

\subsubsection{Plots.} \label{parameters} In the general case the total spectrum
is given by equations (\ref{spectrum})--\eref{K}. In figures 1, 2 and 3 we
compare the spectrum predicted by the usual model (dashed lines), in which only
the absorption/emission channel is present, with the spectrum predicted by our
modified model (solid lines), in which both the absorption/emission channel and
the direct scattering channel are present. In both models we take $\gamma=0.6$,
$\Omega^2=28$; the usual model is characterized by $\delta_\pm=0$,
$\|g_\pm\|^2=0$, $\varepsilon=0$, while as an example of modified model we
choose $\delta_+=-0.03$, $\delta_-=0.13$, $\|g_+\|^2=0.0045$,
$\|g_-\|^2=0.0055$, $\langle g_+| g_- \rangle = -0.004 +\rmi \times 0.002$,
$\varepsilon=-0.001$. We consider three laser detunings $z=0, \ 2.5, \ -2.5$
and four laser bandwidths $y=0,\ 0.5, \ 1, \ 4$. Our choice of the parameters
for the modified model is such that for a monochromatic laser in resonance
(figure 1(a)) the modified spectrum is not quantitatively too different from
the usual one, but its asymmetry is clear. The differences between the two
cases are enhanced by the presence of both the detuning and the bandwidth.

\begin{figure}
\begin{center}
\includegraphics*{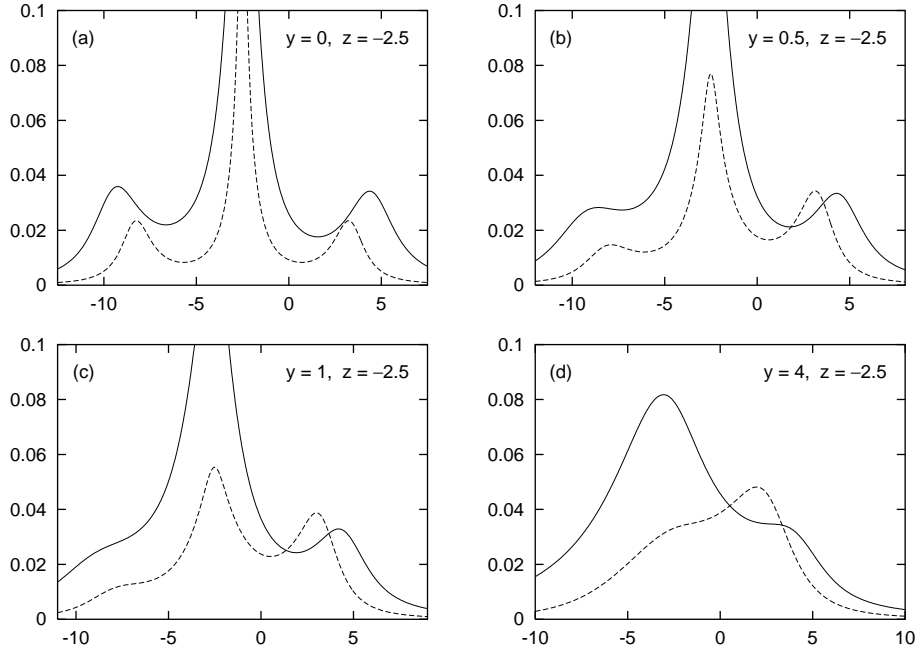}
\end{center}
\caption{The spectrum $\Sigma(x)$ for $\gamma=0.6$, $\Omega^2 = 28$, $z=-2.5$
and  laser bandwidths $y=0,\ 0.5,\ 1, \ 4$. Solid lines: modified model with
the parameters given in section \ref{parameters}; dashed lines: usual model.}
\label{fig3}
\end{figure}

Finally in figure 4 we consider a large bandwidth; however, $y$ is
not much larger than $\Omega^2$, so that we are not exactly in the
case of section \ref{broadband}. Anyway, only a peak survives and
the spectrum is nearly independent from the detuning. In the usual
case we see a nearly Lorentzian shape, centered at $x=0$ ($\nu$
near the atomic frequency $\omega_0$) and with a very weak
residual $\Omega$-dependence. In the modified case the spectrum is
strongly asymmetric and the position of the peak and the width are
$\Omega$-dependent.

\begin{figure}
\begin{center}
\includegraphics*{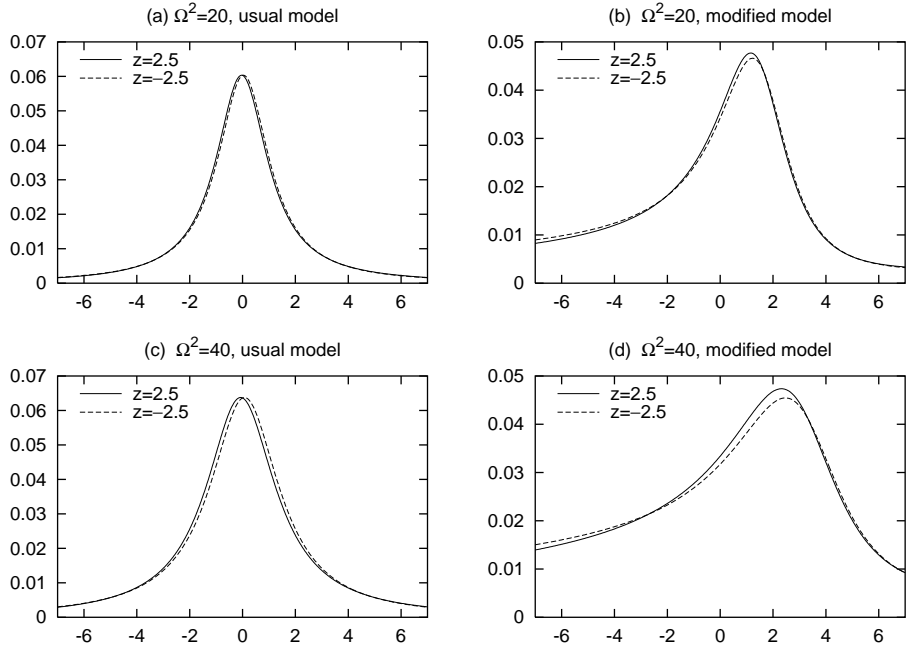}
\end{center}
\caption{The spectrum $\Sigma(x)$ for laser bandwidth $y=50$ and
$\gamma=0.6$, $\Omega^2 = 20,\ 40$, $z=2.5,\ -2.5$. The parameters
of the modified model are given in section \ref{parameters}.}
\label{fig4}
\end{figure}


\section*{References}
\begin{thebibliography}{99}

\bibitem{1HP84}
Hudson R L and Parthasarathy K R 1984 {\it Commun. Math. Phys.} {\bf 93}
301--23

\bibitem{GarC85}
Gardiner C W and Collet M J 1985 {\it Phys. Rev. A} {\bf 31} 3761--74

\bibitem{Gar}
Gardiner C W 1991 {\it Quantum Noise} (Berlin: Springer)

\bibitem{Partha92}
Parthasarathy K R 1992 {\it An Introduction to Quantum Stochastic Calculus}
(Basel: Birkh{\"a}user)

\bibitem{Gar86}
Gardiner C W 1986 {\it Phys. Rev. Lett.} {\bf 56} 1917--20

\bibitem{Bar87}
Barchielli A 1987 {\it J. Phys. A: Math. Gen.} {\bf 20} 6341--55

\bibitem{KW88}
Kennedy T and Walls D F 1988 {\it Phys. Rev. A} {\bf 37} 152--7

\bibitem{AMW88}
Alsing P, Milburn G J and Walls D F 1988 {\it Phys. Rev. A} {\bf 37} 2970--8

\bibitem{LRW88}
Lane A S, Reid M D and Walls D F 1988 {\it Phys. Rev. A} {\bf 38} 788--99

\bibitem{MRW88}
Marte M A, Ritsch H and Walls D F 1988 {\it Phys. Rev. A} {\bf 38} 3577--88

\bibitem{CW88}
Collet M J and Walls D F 1988 {\it Phys. Rev. Lett.} {\bf 61} 2442--4

\bibitem{Bar90}
Barchielli A 1990 {\it Quantum Opt.} {\bf 2} 423--41

\bibitem{WM94}
Wiseman H M and Milburn G J 1994 {\it Phys. Rev. A} {\bf 49} 4110--25

\bibitem{BPag}
Barchielli A and Paganoni A M 1996 {\it Quantum Semiclass. Opt.} {\bf 8}
133--56

\bibitem{Moll}
Mollow B R 1969 {\it Phys. Rev.} {\bf 188} 1969--75

\bibitem{kim-man1}
Kimble H J and Mandel L 1977 {\it Phys. Rev. A} {\bf 15} 689--99

\bibitem{kim-man2}
Kimble H J and Mandel L 1978  {\it Multiphoton Processes} edited by Eberly J H
and Lambropoulos P (New York: Wiley) p 119--28

\bibitem{BarL98}
Barchielli A and Lupieri G  1998 {\it Quantum Probability, Banach Center
Publications, Vol.\ 43} edited by Alicki R, Bozejko M and Majewski W A
(Warsawa: Polish Academy of Sciences, Institute of Mathematics) p 53--62

\bibitem{BarL2000}
Barchielli A and Lupieri G 2000 \JMP {\bf 41} 7181--205

\bibitem{Pero}
Pero N 2000 {\it Thesis} Milan Univ., Phys. Dept.

\bibitem{Eze1}
Ezekiel S and Wu F Y 1978 {\it Multiphoton Processes} edited by Eberly J H and
Lambropoulos P (New York: Wiley) p 145--56

\bibitem{Stroud}
Schuda F, Stroud C R Jr and Hercher M 1974 {\it J. Phys. B: Atom. Molec. Phys.}
{\bf 7} L198--202

\bibitem{Walther1}
Harting W, Rasmussen W, Schieder R and Walther H 1976 {\it Z. Physik A} {\bf
278} 205--10

\bibitem{Eze2}
Grove R E, Wu F Y and Ezekiel S 1977 {\it Phys. Rev. A} {\bf 15} 227--33

\bibitem{Walther2}
Cresser J D, H\"ager J, Leuchs G, Rateike M and Walther H 1982 {\it Dissipative
Systems in Quantum Optics, Topics in Current Physics Vol.\ 27}  edited by
Bonifacio R (Berlin: Springer) p 21-59

\bibitem{Yuen}
Yuen H P and Shapiro J H 1978 {\it IEEE Trans. Inf. Theory} {\bf IT-24} 657--68

\bibitem{Messiah}
Messiah A 1970 \textit{Quantum Mechanics, Vol.\ II} (Amsterdam: North-Holland)

\bibitem{SYMM}
Shapiro J H, Yuen H P and Machado Mata J A 1979 {\it IEEE Trans. Inf. Theory}
{\bf IT-25} 179--92

\bibitem {Goslar}
Barchielli A 1993 {\it Classical and Quantum Systems --- Foundations and
Symmetries --- Proceedings of the II International Wigner Symposium} edited by
Doebner H D, Scherer W and Schroeck F Jr (Singapore: World Scientific) p
488--91


\end{thebibliography}
\end{document}